\documentclass[sigconf, screen]{acmart}
\usepackage{amsmath}
\usepackage{color}
\usepackage{array}
\usepackage{breqn}
\usepackage{adjustbox}	
\usepackage{{inputenc}}
\usepackage{balance}
\usepackage{multirow,tabularx}
\usepackage{booktabs,longtable}
\usepackage{hhline}
\usepackage[flushleft]{threeparttable}
\usepackage{algorithm}
\usepackage{algorithmic}
\usepackage{epsfig}
\usepackage{graphicx}
\usepackage{epstopdf}
\usepackage{thmtools}
\usepackage{enumitem}
\usepackage{verbatim}
\usepackage{amsfonts}
\usepackage{hyperref}
\hypersetup{colorlinks=false,linkcolor=blue,urlcolor=blue,citecolor=red}
\usepackage{xcolor}
\usepackage{subcaption}
\usepackage{wrapfig}
\usepackage{makecell}
\usepackage{shortcuts}
\usepackage{url}
\usepackage{xcolor}
\usepackage{todonotes}
\usepackage{xspace}
\usepackage{tcolorbox}
\usepackage{amsmath}

\newcommand{\codename}{\textsc{MaskDroid}\xspace}

\newcommand{\Lapl}{\mathbf{\mathop{\mathcal{L}}}}

\newcommand{\ie}{\emph{i.e., }}
\newcommand{\eg}{\emph{e.g., }}

\newcommand{\cmmnt}[1]{}
\newtcolorbox{conclusionbox}{colback=gray!8,colframe=black,width=\linewidth,arc=0.6mm, boxrule=0.4pt, left=1mm,right=1mm,top=1mm,bottom=1mm}

\AtBeginDocument{%
  \providecommand\BibTeX{{%
    \normalfont B\kern-0.5em{\scshape i\kern-0.25em b}\kern-0.8em\TeX}}}

\copyrightyear{2024} 
\acmYear{2024}
\setcopyright{rightsretained}
\acmConference[ASE '24]{IEEE/ACM Automated Software Engineering Conference}{27 October - 1 November, 2024}{Sacramento, CA, USA}
\acmDOI{10.1145/3691620.3695008}
\acmISBN{979-8-4007-1248-7/24/10}

\title{\codename: Robust Android Malware Detection with Masked Graph Representations}

\author{Jingnan Zheng{$^*$}}
\affiliation{%
  \institution{National University of Singapore}
  \country{Singapore}}
\email{jingnan.zheng@u.nus.edu}

\author{Jiaohao Liu{$^*$}}
\affiliation{%
  \institution{National University of Singapore}
  \country{Singapore}}
\email{jiahao99@comp.nus.edu.sg}

\author{An Zhang{$^\dagger$}}
\affiliation{%
  \institution{National University of Singapore}
  \country{Singapore}}
\email{anzhang@u.nus.edu}

\author{Jun Zeng}
\affiliation{%
  \institution{National University of Singapore}
  \country{Singapore}}
\email{junzeng@u.nus.edu}

\author{Ziqi Yang{$^\S$}}
\affiliation{%
  \institution{Zhejing University}
  \country{China}}
\email{yangziqi@zju.edu.cn}

\author{Zhenkai Liang}
\affiliation{%
  \institution{National University of Singapore}
  \country{Singapore}}
\email{liangzk@comp.nus.edu.sg}

\author{Tat-Seng Chua}
\affiliation{%
  \institution{National University of Singapore}
  \country{Singapore}}
\email{chuats@comp.nus.edu.sg}

\thanks{$^*$ Jingnan Zheng and Jiaohao Liu contribute equally to this work}
\thanks{$^\dagger$ An Zhang is the corresponding author of this work.}
\thanks{$^\S$ Ziqi Yang is affiliated with the State Key Laboratory of Blockchain and Security and the Hangzhou High-Tech Zone (Binjiang) Institute of Blockchain and Data Security.}

\acmConference[ASE]{IEEE/ACM Automated Software Engineering Conference}{2024}{Sacramento, CA, US}
\acmYear{2024}

\begin{document}

\begin{abstract}
Android malware attacks have posed a severe threat to mobile users, necessitating a significant demand for the automated detection system.
Among the various tools employed in malware detection, graph representations (\eg function call graphs) have played a pivotal role in characterizing the behaviors of Android apps.
However, though achieving impressive performance in malware detection, current state-of-the-art graph-based malware detectors are vulnerable to adversarial examples.
These adversarial examples are meticulously crafted by introducing specific perturbations to normal malicious inputs.
To defend against adversarial attacks, existing defensive mechanisms are typically supplementary additions to detectors and exhibit significant limitations, often relying on prior knowledge of adversarial examples and failing to defend against unseen types of attacks effectively.

In this paper, we propose \codename, a powerful detector with a strong discriminative ability to identify malware and remarkable robustness against adversarial attacks.
Specifically, we introduce a masking mechanism into the Graph Neural Network (GNN) based framework, forcing \codename to recover the whole input graph using a small portion (\eg 20\%) of randomly selected nodes.
This strategy enables the model to understand the malicious semantics and learn more stable representations, enhancing its robustness against adversarial attacks.
While capturing stable malicious semantics in the form of dependencies inside the graph structures, we further employ a contrastive module to encourage \codename to learn more compact representations for both the benign and malicious classes to boost its discriminative power in detecting malware from benign apps and adversarial examples.
Extensive experiments validate the robustness of \codename against various adversarial attacks, showcasing its effectiveness in detecting malware in real-world scenarios comparable to state-of-the-art approaches.

\end{abstract}


\maketitle

\section{Introduction}
\label{sec:introduction}
Android, as one of the most prevalent smartphone operating systems, has dominated over 85\% of the mobile OS market share since 2018~\cite{li2018significant}. 
However, its popularity and open nature have also made it a primary target for cyberattacks~\cite{liu2022deep, judy}. 
For example, Android permits the installation of applications from unverified sources, such as third-party markets, thereby providing attackers with easy means to bundle and distribute malware-infected apps~\cite{arp2014drebin}. 
Android malware, a.k.a., malicious software, has become one of the primary security threats to the Android platform, with the number of malware samples increasing exponentially over the years~\cite{harmful}.

To mitigate these threats, machine learning (ML) has been widely adopted to automatically extract malicious patterns from various APK features for Android malware detection~\cite{gao2024comprehensive,liu2024unraveling,liu2022deep}.
According to the features used, there are two research lines: syntax-based~\cite{arp2014drebin,li2021robust,wu2021android,aafer2013droidapiminer} and semantic-based detectors~\cite{mamadroid,wu2021homdroid,hou2017hindroid,wu2019malscan,he2022msdroid}.
Syntax-based methods typically utilize discrete features such as permissions, API calls, and intents to model app behaviors.
However, these methods might overlook the underlying program semantics, limiting their detection capability~\cite{wu2021homdroid,he2022msdroid}.
To address this issue, a notable trend is to distill program semantics from apps' graph representations for malware detection.
Among these graph-based approaches, Function Call Graphs (FCGs) and their variants are extensively used and have proven effective~\cite{mamadroid,wu2019malscan,he2022msdroid}, as they encapsulate the invocation relationship among API calls, offering deep insights about how an app works~\cite{li2023black}.
For example, MsDroid~\cite{he2022msdroid} utilizes code snippets around sensitive API calls in FCGs to model app behaviors and leverages graph neural networks (GNNs) to capture corresponding semantics for malware detection.

While existing graph-based malware detectors have demonstrated promising results in detecting malware, they are vulnerable to adversarial examples crafted by carefully introducing perturbations to malicious inputs~\cite{chen2019android, zhao2021structural, zhang2022semantics,gao2024comprehensive,liu2024unraveling}.
For example, adversaries can modify the most influential edges or nodes in apps' graph representations to evade the decision boundary, thereby undermining the detectors' performance~\cite{li2023black,zhao2021structural}.
This is because of the inherent fragility of ML models, where small but intentional perturbations can lead to incorrect predictions with high confidence~\cite{goodfellow2014, croce2020reliable, madry2017towards, athalye2018obfuscated}.
Unfortunately, there are few defense mechanisms available to enhance the robustness of graph-based malware detectors against adversarial attacks, except for several general-purpose supplemental strategies, like adversarial training~\cite{goodfellow2014, al2018adversarial, wang2017adversary}.
However, these strategies often require a large number of adversarial examples for training, which is impractical in real-world scenarios~\cite{gu2014towards, ororbia2016unifying}.
Additionally, they may not be effective against unseen types of adversarial examples.
As such, we argue that the current defense mechanisms are insufficient to meet the demands posed by adversarial attacks, highlighting the need for more robust solutions, especially those that enhance the detector itself.

In this paper, we aim to propose a novel graph-based Android malware detector, \codename, that can effectively handle adversarial attacks while maintaining comparable detection accuracy.
To achieve this goal, we need to learn stable representations of malicious behaviors that remain consistent even if the input graph is adversarially perturbed.
Specifically, malware developers often conceal their intentions by mimicking the behavior of benign apps, embedding a small portion of malicious code within a bulk of benign processes~\cite{medium, hou2017hindroid}.
When representing the app in a graph form, the malicious code appears as a small sub-graph embedded within the overall graph representation.
Adversarial examples preserve the codes (\ie malicious subgraphs) responsible for the malicious operations, while strategically introducing additional nodes or edges to confound the detectors~\cite{li2021robust}.
Based on this observation, we note that if we can learn a stable representation encoding the malicious behavior, we can achieve a robust detector that can effectively detect both malware and adversarial examples.

To guide \codename to learn stable representations of malicious behaviors within the graph representations, our methodology involves introducing uncertainty and forcing the model to contend with it through a \textit{reconstruction module}.
Specifically, we first apply a random masking mechanism~\cite{he2022masked,vincent2008extracting} to the input graph, where a substantial proportion of nodes (\eg 80\%) are masked out.
Then we use graph neural networks to encode and decode the masked graph, forcing \codename to recover the masked node features using the remaining nodes (\eg 20\%).
By doing so, \codename gains a holistic understanding of the malicious behaviors and develops the ability to generate stable representations even when the input is perturbed, enhancing its robustness against adversarial attacks.

With stable representations that comprehensively grasp the semantics of the malicious structures, we proceed to determine the class (\ie benign or malicious) of the input graph.
Instead of performing direct classification, \codename incorporates a \textit{contrastive module}~\cite{yao2022pcl} to enhance its discriminative power.
The insight behind this is that samples within the same class can complement each other and should be pulled closer together, while those in different classes should be pushed apart.
To this end, we define two proxy representations as anchors for the benign and malicious classes.
During the training process, we calculate the distance between the input instance and the two proxies to update their positions.
The prediction is made by checking which proxy the instance is closer to.
The contrastive module further compresses the representations and refines the decision boundary between benign and malicious classes, enhancing \codename's ability to distinguish malware from benign apps and adversarial examples.


To investigate the effectiveness and robustness of \codename in Android malware detection, we conduct extensive experiments on a comprehensive dataset comprising 102,459 benign and 11,751 malicious apps collected over five years, from 2016 to 2020.
We further compare \codename with five state-of-the-art malware detectors, including three graph-based detectors, \ie MamaDroid~\cite{mamadroid}, MalScan~\cite{wu2019malscan}, and MsDroid~\cite{he2022msdroid}, and two syntax-based detectors, \ie Drebin~\cite{arp2014drebin} and RAMDA~\cite{li2021robust}.
Experimental results demonstrate that \codename achieves the most robust performance against adversarial attacks under various settings (e.g., reducing the attack success rate from 41.54\% to 32.0\%) while maintaining comparable detection accuracy to the state-of-the-art detectors.
Through ablation studies, we further validate that each design choice (e.g., contrastive module and reconstruction module) in \codename contributes to its robustness and effectiveness.

In summary, we make the following contributions:


\begin{itemize}[leftmargin=10pt]
    \item To the best of our knowledge, we are the first to enhance the robustness of the graph-based malware detection model against adversarial attacks without sacrificing detection accuracy.
    \item We present \codename, a novel graph-based Android malware detector that utilizes a reconstruction task to guide the model to learn stable representations of malicious behaviors, while incorporating a contrastive module to enhance its discriminative power for malware detection.
    \item We conduct extensive evaluations against five state-of-the-art approaches on data sourced from AndroZoo ~\cite{allix2016androzoo}. Experimental results show that \codename exhibits superior robustness against adversarial attacks compared to the baselines while maintaining promising detection accuracy.
    Our codes are available at \url{https://github.com/SophieZheng998/MaskDroid}.
\end{itemize}

\section{Preliminaries}
\label{sec:prelimimary}
In this section, we first introduce commonly used graph representations (\ie Function Call Graph and its variants) in Android malware detection.
Then, we formally formulate the problem of graph-based malware detection and adversarial example attacks.

\begin{figure}[t]
    \centering
    \includegraphics[width=0.87\linewidth]{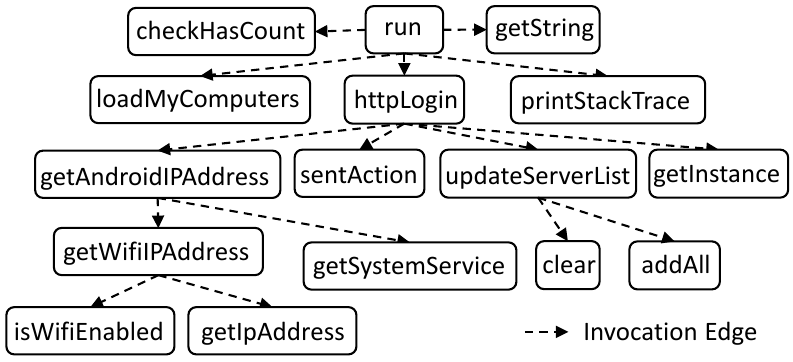}
    \vspace{-0.2cm}
    \caption{Partial Function Call Graph (FCG) of an app.}
    \vspace{-0.4cm}
    \label{fig:cg}
\end{figure}

\subsection{Graph Representations}
The graph representations of Android apps encode both the semantic and structural information and have been widely used in Android malware detection~\cite{hou2017hindroid,mamadroid,wu2019malscan,he2022msdroid,he2022msdroid}.
Among these, the Function Call Graph (FCG) is a popular representation that captures the caller-callee relationships among the API calls in an app.
Figure~\ref{fig:cg} depicts a partial Function Call Graph (FCG) of a real-world Android application~\cite{appexample}, where nodes (\eg \textit{getWifiIPAddress()}) represent API calls, and the edges denote method invocations (\eg \textit{updateServerList()} calls \textit{addALL()}).
Detectors like MalScan~\cite{wu2019malscan} and HomDroid~\cite{wu2021homdroid} analyze FCGs akin to social networks, leveraging centrality and community detection algorithms to uncover malicious patterns for malware detection.

Additionally, several variants of FCGs have been proposed to model app behaviors.
For example, MamaDroid~\cite{mamadroid} abstracts the nodes of FCGs according to their packages or family names to construct a higher-level, abstracted graph representation.
MsDroid~\cite{he2022msdroid} uses sensitive API calls (\textit{e.g., getSystemService()}) as seed nodes to generate graph snippets around them, modeling apps as a collection of subgraphs.
This is because sensitive behaviors are often carried out by a small proportion of the code requiring the invocation of sensitive API calls to achieve their goals.
MsDroid further utilizes the opcode and required permissions of functions as the node features to initialize the graph representation.
In our study, we adopt the graph structure proposed in MsDroid due to its simplicity and proven effectiveness in detecting Android malware.

\subsection{Problem Formulation}
\label{sec:formulation}
\noindent \textbf{Android Malware Detection.}   
Graph-based Android malware detectors take the graph representation of an app as input and output the probability of it being malicious.
Here, we formally define the input graph as $\mathcal{G} = (\mathcal{V}, \mathcal{E}, \mathcal{X})$, where each node $v \in \mathcal{V}$ represents an API call, and each edge $e_{(u,v)} \in \mathcal{E}$ denotes the invocation from node $u$ to node $v$.
The set $\mathcal{X}$ collects the features of the nodes.
The goal of the malware detector is to learn a classifier $\mathcal{C}: \mathcal{G} \rightarrow \{0, 1\}$, where $0$ denotes a benign app, and $1$ denotes malware.

\vspace{0.1cm}
\noindent \textbf{Adversarial Examples.}
Adversarial examples are crafted to mislead learning classifiers by introducing perturbations to the input features while preserving the malicious functionalities~\cite{li2023black,zhao2021structural,pierazzi2020intriguing,chen2019android,li2020adversarial}.
In the context of graph-based Android malware detection, these adversarial examples can be represented as deliberately altered target graph-based features to bypass the classifier $\mathcal{C}$.
Suppose $\mathcal{P}$ denotes the perturbation operation, and $L(\cdot)$ is the label predicted by the classifier $\mathcal{C}$. 
Then, the adversarial example $\mathcal{G}'$ can be formulated as $\mathcal{G}' = \mathcal{P}(\mathcal{G}) = \mathcal{G} + \delta$, where $\delta$ signifies the perturbations to the graph structure, such as adding nodes and edges.
The adversarial manipulation process can be represented as:
\begin{gather}
    L(\mathcal{C}(\mathcal{G})) \neq L(\mathcal{C}(\mathcal{P}(\mathcal{G}))).
\end{gather}

Intuitively, an ideal Android malware detector should be highly effective in identifying malware while also being robust against adversarial examples.
However, current models predominantly emphasize detection effectiveness, often at the expense of robustness~\cite{wu2019malscan,li2023black,he2023efficient,arp2014drebin}.
Only a few studies have attempted to improve the robustness of Android malware detectors against adversarial examples.
For instance, RAMDA~\cite{li2021robust} is a state-of-the-art approach that aims to improve detectors' robustness by squeezing the room of adversarial examples in the latent space.
Nonetheless, this method sacrifices detection effectiveness since it also filters out benign apps that are close to adversarial examples.
As such, designing robust and effective Android malware detectors remains an open challenge~\cite{liu2024unraveling,gao2024comprehensive}.
In this study, we propose a novel graph-based approach, \codename, detailed in Section~\ref{sec:method}, to further advance this field.
Our approach not only bolsters robustness against adversarial examples but also ensures high detection effectiveness through a more precise interpretation of the program semantics of apps.

\section{Methodology}
\label{sec:method}
In this section, we present a novel learning framework, \codename, designed to improve the robustness of Android malware detection without compromising detection performance.
Guided by a masking mechanism, \codename can more effectively explore the structural and semantic information encoded in the graph representations, thereby enhancing the understanding of potential malicious behaviors.
Additionally, we further incorporate a proxy-based contrastive learning module to boost \codename's ability to discriminate between benign and malicious instances.

\subsection{Overview}
\label{sec:overview}
\begin{figure*}[t]
    \centering
    \includegraphics[width=0.86\linewidth]{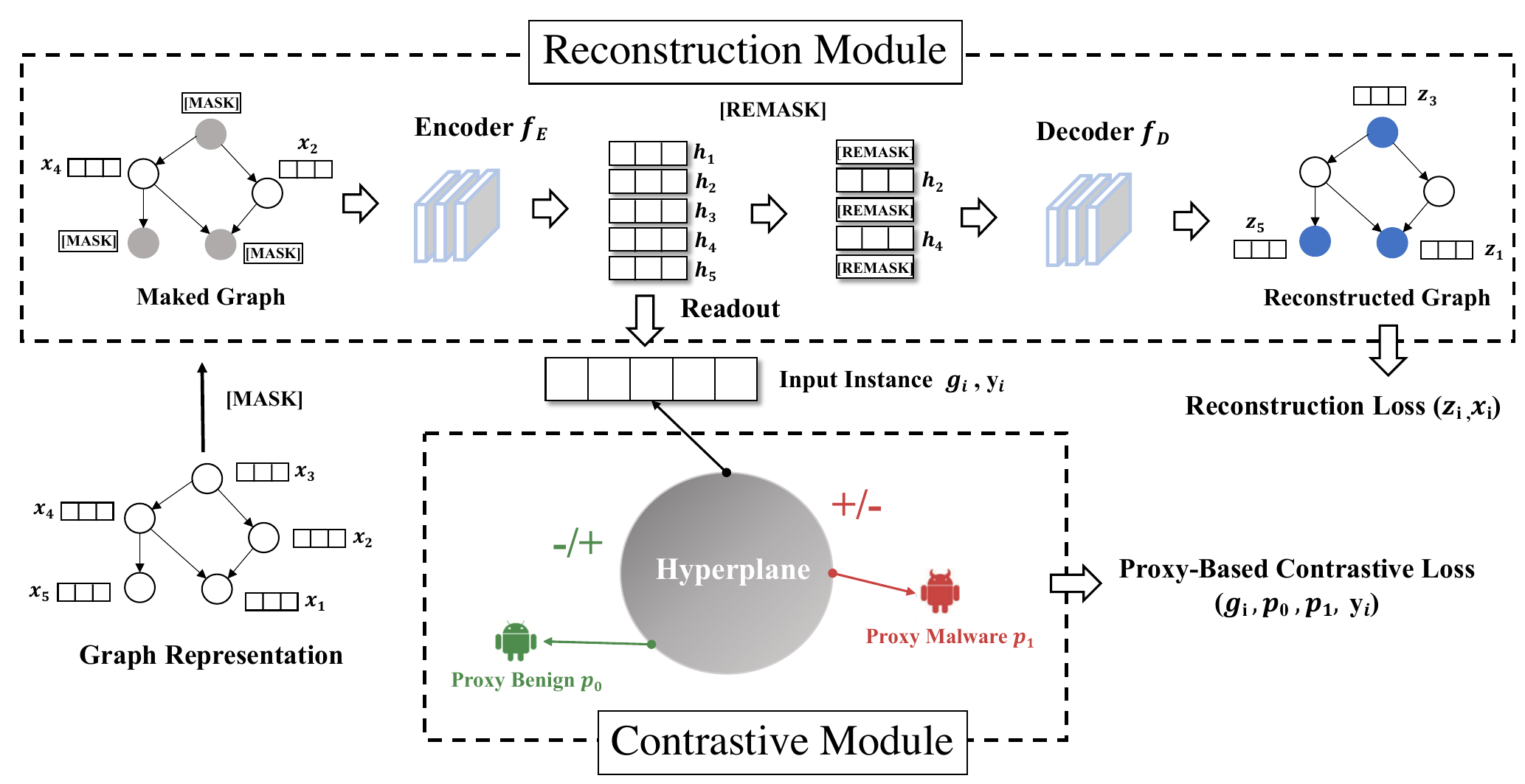}
    \vspace{-0.2cm}
    \caption{The framework of \codename. The training phase comprises two modules. The upper dashed-line bracket represents the self-supervised reconstruction module, while the lower bracket represents the proxy-based contrastive learning module.}
    \vspace{-0.3cm}
    \label{fig:framework}
\end{figure*}

Figure~\ref{fig:framework} illustrates the high-level overview of \codename.
The model consists of two main components: (1) a self-supervised reconstruction module that utilizes graph neural networks (GNNs) along with a graph mask mechanism to learn semantics of the input graphs, and (2) a proxy-based contrastive learning module, which leverages the mutual information across samples in similar and dissimilar classes to enhance the model's discriminatory power.

In the reconstruction part, we first mask a proportion of the nodes in the input graph.
A GNN encoder is then applied to map the features of each node into a latent space. 
Following this, a GNN decoder reconstructs the masked-out nodes based on the latent representations of the remaining nodes.
This self-supervised task promotes \textit{the learning of the underlying structures and dependencies within the input graphs, thereby deepening the model's understanding of app behaviors.}

With the graph-level representation obtained from the graph encoder, we proceed to the contrastive module.
The principle behind this module is that \textit{apps within the same class (i.e., benign or malicious) should be closer to each other, while apps from different classes should be more distant.}
To achieve this, we initialize two proxy representations as the anchors of benign and malicious classes, respectively.
During the training phase, we pull a sample closer to the anchor of its class and push it away from the anchor of a different class.
These proxy representations are updated simultaneously to maintain their roles as class anchors.

To predict the category of an app, \codename disables the mask mechanism and processes the input instance through the encoder to obtain a graph-level representation.
Finally, it determines whether the app is benign or malicious based on which proxy the graph-level representation is closer to.

\subsection{Reconstruction Module}
\label{sec:reconstruction}
We now present the details of the self-supervised reconstruction module in \codename.
For better understanding, we will first recap the graph representation used in \codename before delving into the module itself.

\bulletpoint{Recap of Graph Representation}
The graph representation utilizes the function call graph (FCG) as its input.
Specifically, we begin by identifying a set of sensitive API calls (\textit{e.g., getIpAddress()}) within an app's FCG to serve as seed nodes.
Next, we extract subgraphs centered on these seed nodes at a fixed depth to form the input graph.
It is worth noting that focusing on sensitive API calls and their surrounding contexts can effectively capture apps' malicious behaviors, as these behaviors are often carried out by a small part around sensitive API calls.
This strategy has been validated in previous studies~\cite{wu2021homdroid,wu2019malscan,gu2024gsedroid}.
To further enhance the representation, each node is initialized with the opcode and required permissions associated with its corresponding API call.
The opcode and permissions are pivotal in understanding the semantics of API calls, as they provide critical insights into app behaviors~\cite{he2022msdroid,kim2018multimodal}.
For instance, the opcode \textit{invoke} signifies that one function must call another to complete a task, while the permission \textit{SEND_SMS} denotes the app's capability to send text messages.

Here, we begin with the initialization of node features and then introduce how we mask and reconstruct the graph to learn the underlying program semantics in a self-supervised manner.

\subsubsection{Node Initialization}
\label{sec:masking}
Given the graph representation, $\mathcal{G} = (\mathcal{V}, \mathcal{E}, \mathcal{X})$, each node $v \in \mathcal{V}$ is characterized by its attributes (\ie opcode and permissions) that describe the API call.
To capture this information, we initialize the node feature $\textbf{x}_v \in \mathbb{R}^d$ as the concatenation of the embeddings for the opcode and permissions:
\begin{gather}
    \textbf{x}_v =  n_{v_{op}}\ ||\ n_{v_{per}},
\end{gather}
where $n_{v_{op}}$ and $n_{v_{per}}$ represent the one-hot encodings of the opcode and permissions, respectively.

\subsubsection{Graph Reconstruction}
\label{sec:graph_reconstruction}
Our objective is to learn a high-quality representation that is resilient to adversarial attacks while ensuring optimal detection performance.
The reconstruction module supports this goal by encouraging the model to recover masked nodes using unmasked nodes, effectively capturing the underlying structural and semantic information within the input graph.
Consequently, even if the input graph is partially corrupted, \codename retains its ability to discern malicious semantics, making it more robust to adversarial attacks.

\bulletpoint{Graph Masking and Encoder}
We apply uniform random sampling to choose a subset of nodes $\widetilde{\mathcal{V}} \subset \mathcal{V}$ and mask their corresponding representations with a learnable vector $\textbf{x}_{[M]}$.
This strategy ensures that for each node, its neighbors are neither all masked nor all visible~\cite{hou2022graphmae}.
Based on this, it is easier to recover the masked nodes with their neighboring unmasked nodes, facilitating the training of the model to understand the graph structure.
Formally, the node feature $\widetilde{\textbf{x}}_v$ of the masked graph $\widetilde{\mathcal{G}}$ can be defined as:
\begin{gather}
    \widetilde{\textbf{x}}_v = 
    \begin{cases} 
    \textbf{x}_{[M]} &   v \in \widetilde{\mathcal{V}},  \\
    \textbf{x}_v &  v  \notin \widetilde{\mathcal{V}}.
    \end{cases}
\end{gather}  
Considering the inherent graph nature of the masked graph, \codename utilizes graph neural networks~\cite{hamilton2017inductive} (GNNs) to analyze the structural information and learn the corresponding node representations, \ie embeddings.
GNNs are particularly suited for this as they recursively propagate and aggregate node features across edges, enabling the model to capture both local and global graph dependencies.
Take the sequence (\textit{getAndroidIPAddress() $\rightarrow$ getWifiIPAddress() $\rightarrow$ isWifiEnabled() $|$ getIpAddress()}) depicted in Figure~\ref{fig:cg} as an example.
If we examine the nodes in isolation, we only see that the app invokes several Wifi-related functions, \textit{e.g., getWifiIPAddress()}.
It is difficult to capture how the app executes the entire process of checking the network status and obtaining the IP address.
However, by leveraging GNNs, \codename can better explore the multi-hop relation between nodes --- such as propagating information from \textit{getWifiIPAddress()} to \textit{isWifiEnabled()} and \textit{getIpAddress()} --- enhancing the understanding of how the app works.
Formally, the representation of a node $v$ at layer $l+1$ is updated by aggregating the embeddings of its neighbors as follows:
\begin{align}
    \widetilde{\textbf{x}}_v^{(l+1)} = \sigma((\widetilde{\textbf{x}}_v^{(l)} + \sum _{u \in \mathcal{N}_{v}} \frac{\widetilde{\textbf{x}}_{v}^{(l)}}{\sqrt{\left | \mathcal{N}_{u} \right | \left | \mathcal{N}_v\right |}})\textbf{W}_{\alpha}^{(l)}) \label{eq:gnn},
\end{align}
where $\mathcal{N}_{v}$ represents the set of neighbors of node $v$, $\textbf{W}_{\alpha}^{(l)}$ is the weight matrix at layer $l$, and $\sigma$ is the activation function, such as ReLU or LeakyReLU.
After $L$ layers of propagation, the final node embeddings are obtained as $\textbf{h}_v = \widetilde{\textbf{x}}_v^{(L)}$.

\bulletpoint{Graph Remasking and Decoder}
With the latent representation $\textbf{h}_v$ for each node in the masked graph, the next step is to recover the masked nodes and reconstruct the original graph.
If \codename can accurately recover the masked nodes, it indicates that the model can infer the missing information based on the surrounding nodes, enhancing its robustness to adversarial attacks.
For example, in Figure~\ref{fig:cg}, \textit{getWifiIPAddress()} is masked out, and the model is trained to recover it based on its surrounding nodes, such as \textit{getAndroidIPAddress()}, \textit{isWifiEnabled() and getIpAddress()}.
This implies that even if an adversary partially alters the graph, \codename can still deduce that the app is attempting to obtain the Wifi status and IP address.
With this stable understanding, \codename can still make accurate predictions.

To boost the model's ability to recover masked nodes based solely on their surrounding nodes, we re-mask the masked nodes in $\widetilde{\mathcal{V}}$ to prevent the model from memorizing them~\cite{hou2022graphmae}.
Specifically, the re-masked representation $\widetilde{\textbf{h}}_v$ is defined as follows:
\begin{gather}
\widetilde{\textbf{h}}_v = 
\begin{cases} 
\textbf{h}_{[M]} & v \in \widetilde{\mathcal{V}},  \\
\textbf{h}_v &  v  \notin \widetilde{\mathcal{V}}. 
\end{cases}
\end{gather}
where $\textbf{h}_{[M]}$ is a learnable vector used to re-mask selected nodes.
In our implementation, we simply set the re-masked vector $\textbf{h}_{[M]}$ to zero due to its effectiveness, leaving more sophisticated strategies for future work.
The re-masked representation is then fed into the decoder $f_D$ to reconstruct the original node features.
Similar to the encoder, the decoder also utilizes the GNN architecture, which is better to capture the structural information and reconstruct the masked nodes.
\begin{gather}
    \textbf{z}_v = f_D(\widetilde{\textbf{h}}_v) = \text{GNN}(\widetilde{\textbf{h}}_v).
\end{gather}
Here, for clarity, we omit the details of how the decoder propagates and aggregates information along the graph, as this process can be designed in a manner similar to the encoder.

To train the model, we define the reconstruction loss $\Lapl_{\text{rec}}$ to measure the discrepancy between the original node features and the reconstructed features.
Particularly, we employ the cosine similarity to measure their distance as follows:
\begin{gather}
    \Lapl_{\text{rec}} = \frac{1}{|\widetilde{\mathcal{V}}|}
    \sum_{v\in\widetilde{\mathcal{V}}} (1 - \frac{\textbf{x}^{T}_{v}\textbf{z}_v}{\rVert \textbf{x}_v \rVert \cdot \rVert \textbf{z}_v \rVert })^{2}.
\end{gather}

In summary, through the self-supervised graph reconstruction learning task, \codename gains a holistic understanding of the graph structures and semantics, which is crucial for the success of the subsequent discrimination task.

\subsection{Contrastive Module}
\label{sec:contrastive}
With the stable representation that captures app semantics, we now turn to the contrastive module, which aims to enhance the model's discriminatory power.
The principle behind this module is intuitive: apps executing similar behaviors should cluster together and mutually reinforce each other, while apps from different classes should be more distant from each other.

Towards this end, we adopt a proxy-based contrastive learning strategy~\cite{yao2022pcl} to explore the mutual information across samples in similar and dissimilar classes.
Since Android malware detection is a binary classification task, we define two proxies, $\textbf{p}_0$ and $\textbf{p}_1$ for benign and malicious classes, respectively.
The proxy representations are learnable vectors with random initialization, utilized to capture the distinctions between benign and malicious apps, facilitating their separation in the latent space. 
During training, each instance is pulled closer to the proxy of its own class while being pushed further from the other proxy. 
The two proxy vectors are updated simultaneously with the model parameters throughout the training process to maintain their roles as class anchors.
Assume the graph-level representation obtained from the encoder $f_E$ is $\textbf{g}_i$ for the $i$-th instance.
Each instance has a supervised label $y_i$ indicating whether it belongs to the benign or malicious class, where 0 represents benign and 1 represents malicious.
The contrastive learning process can be defined as follows:
\begin{align}
    \Lapl_{\text{cl}}  
    &=   y_i \cdot
   \left[( \frac{\textbf{g}_i \cdot  \mathbf{p_0}}{\|\textbf{g}_i\| \|\mathbf{p_0}\| })^{2} +
   (1- \frac{\textbf{g}_i \cdot  \mathbf{p_1}}{\|\textbf{g}_i\| \|\mathbf{p_1}\| }  )^{2}\right] \\
   &+  (1-y_i) \cdot \left[( \frac{\textbf{g}_i\cdot  \mathbf{p_1}}{\|\textbf{g}_i\| \|\mathbf{p_1}\| })^{2} +
   (1- \frac{\textbf{g}_i \cdot  \mathbf{p_0}}{\|\textbf{g}_i\| \|\mathbf{p_0}\| }  )^{2}\right] \notag.
\end{align}

After the training phase, the two proxies $\textbf{p}_0$ and $\textbf{p}_1$ aggregate the information of all instances in or not in their respective classes, serving as the class anchors for the inference phase.




\subsection{Android Malware Detection}
\label{sec:detector}
To optimize the model for Android malware detection, we combine the reconstruction and contrastive modules into a joint training framework.
The final objective of \codename is defined as:
\begin{gather}
    \Lapl=\lambda_{1}\cdot\Lapl_{\text{rec}}+\lambda_{2}\cdot\Lapl_{\text{cl}},
\end{gather}
where $\lambda_1$ and $\lambda_2$ are the hyper-parameters to control the strength of two modules.
By minimizing this objective, \codename learns a high-quality representation that captures app semantics and maintains two anchors that consider all instances in the training set, enhancing the model's robustness and discriminatory power.

In the detection phase, \codename first transforms the input graph into a graph-level representation using the encoder $f_E$.
Then it calculates the distance between the graph-level representation and the two anchors to determine whether the input instance is benign or malicious.
 
\section{Evaluation}
\label{sec:evaluation}
In this section, we evaluate the performance of \codename by answering the following research questions (RQs):
\begin{itemize}[leftmargin=10pt]
\item \textbf{RQ1:} Does \codename successfully improve the robustness against different adversarial attacks (\eg white-box and black-box attacks) compared to its baselines?
\item \textbf{RQ2:} Does \codename sacrifice detection effectiveness to enhance its robustness against adversarial attacks?
\item \textbf{RQ3:} To what extent do different design choices affect the performance of \codename on counteracting adversarial attacks and detecting malware?
\item \textbf{RQ4:} Does \codename require more computational resources to complete its detection?
\end{itemize}
\subsection{Experimental Setup}
\label{sec:setup}

\subsubsection{Implementation}
\label{sec:implementation}
We utilize Androguard~\cite{Androguard} to decompile APKs and extract the Function Call Graph (FCG) for each app. 
Based on the extracted FCGs, we construct the input graph for our model in a three-step procedure:
(a) we traverse the FCG to collect sensitive API calls.
It is worth noting that we focus on the sensitive API calls reported by PSCount~\cite{au2012pscout} and Axplorer~\cite{backes2016demystifying}, as they include the most commonly used sensitive API calls in Android malware~\cite{wu2019malscan,wu2021homdroid,he2022msdroid};
(b) we take the collected sensitive API calls as roots and perform a breadth-first search to collect the call sequences within a certain depth (\ie 2), then merge them into the input graph;
(c) we initialize the nodes in the input graph with their corresponding opcodes and permissions.

To find optimal hyper-parameters for \codename, we employ a grid search strategy.
The learning rate is tuned from \{0.1,0.01,0.001, 0.0001\}, and the mask rate is searched within \{0.1,0.2, 0.3, 0.4, 0.5, 0.6,0.7, 0.8,0.9\}.
For experiments in Sections ~\ref{sec:robust} and ~\ref{sec:effect}, we use the entire dataset from 2016 to 2020, finding that mask rate = 0.8 yields the best performance. As such, we choose 0.8 as the masking rate. 
For the ablation study in Sections ~\ref{sec:ablation} and ~\ref{sec:efficiency}, we use data from 2020 and find 0.5 is the optimal masking rate.
Furthermore, the number of GNN layers in both the encoder $f_E$ and decoder $f_D$ is tuned among \{1,2,3\}.
Based on achieving the best performance, we present results under the configuration of a mask rate of 0.8, 0.001 learning rate, and two 2-layer GNNs for encoder and decoder.
In addition, $\lambda1$ and $\lambda2$ are set to be equal in our experimental setting.

All experiments are performed on a server with an Intel Xeon Gold 6248 CPU @ 2.50GHz, 188GB physical memory, and an NVIDIA Tesla V100 GPU. The OS is Ubuntu 20.04.2 LTS.

\subsubsection{Datasets}
\label{sec:dataset}
To rigorously assess \codename's performance, we source our dataset from AndroZoo~\cite{allix2016androzoo}, a continuously expanding repository of Android apps that aggregates apps from several sources, such as Google Play, Appchina, and Anzhi.
Our dataset consists of 114,210 apps, among which 102,459 are benign and 11,751 are malicious, spanning from 2016 to 2020, as depicted in Table~\ref{tab:data_statistics}. 

Importantly, the dataset adheres to the guidance proposed by \textsc{Tesseract}~\cite{pendlebury2019tesseract}.
\textit{Avoid Grayware}: 
The ambiguous nature of grayware can potentially skew the performance of learning models.
To counteract this threat, we utilize the positive anti-virus alerts from VirusTotal~\cite{virustotal}, represented by $p$, to filter out grayware.
In particular, apps with $p \geq 4$ are labeled malicious, whereas those with $p = 0$ are classified as benign.
\textit{Goodware-to-Malware Ratio}:
Previous studies have verified that the ratio of benign to malicious apps in the wild is notably imbalanced, with malware constituting a small fraction (~10\%) \cite{pendlebury2019tesseract, api, continuous}.
To more accurately gauge the effectiveness of our approach in real-world scenarios, we ensure the malware proportion in our dataset is set at 10\%.
Additionally, we sample the dataset from 2016 to 2020 to cover a wide range of apps and reflect the temporal dynamics of malware evolution.
In our experiments, we randomly split each dataset into three disjoint sets: training, validation, and testing, with proportions of 70\%, 20\%, and 10\%, respectively.
We also ensure all the disjoint sets exhibit a 9:1 ratio of benign to malicious apps.

\begin{table}[t]
 \centering
 \caption{Evaluation Dataset Statistics. The samples cover three years from 2016 to 2020 and maintain a 9:1 ratio of benign to malicious apps.}
  \vspace{-10pt}
 \begin{adjustbox}{width=0.89\linewidth, center}
 \begin{tabular}{l|cccc}
\hline
\textbf{Year}& \textbf{Begin} & \textbf{Malware} & \textbf{M+B} & \textbf{M/(M+B)} \\ \hline
\textbf{2016}& 21,292& 2,390& 23,682& 10.1\% \\
\textbf{2017}& 21,006& 2,389& 23,395& 10.2\% \\
\textbf{2018}& 20,099& 2,326& 22,425& 10.4\% \\
\textbf{2019}& 20,260& 2,345& 22,605& 10.4\% \\
\textbf{2020}& 19,802& 2,301& 22,103& 10.4\% \\ \hline
\textbf{Total} & 102,459 & 11,751 & 114,210& 10.3\% \\ \hline
\end{tabular}
 \end{adjustbox}
 \label{tab:data_statistics}
  \vspace{-0.3cm}
\end{table}

\subsubsection{Baselines}
\label{sec:baselines}
To comprehensively investigate the performance of \codename, we compare it with three state-of-the-art graph-based detection approaches: MamaDroid, MalScan, and MsDroid, as well as two non-graph-based detectors: Drebin and RAMDA.

\begin{itemize}[leftmargin=10pt]
 \item \textbf{MamaDroid}~\cite{mamadroid}: This method abstracts Function Call Graphs (FCGs) according to the package or family level of function names. It then constructs Markov Chains and calculates the transition probabilities between different nodes, which serve as the feature vector to train a Random Forest classifier.
 \item \textbf{MalScan}~\cite{wu2019malscan}: This algorithm treats FCGs as social networks and conducts centrality analysis on sensitive API calls to capture APKs' semantics for classification.
 \item \textbf{MsDroid}~\cite{he2022msdroid}: This detector is built on the key insight that malicious operations often in code snippets around sensitive API calls. It thus models app-sensitive behaviors as subgraphs around sensitive API calls and feeds them into a GNN classifier.
 \item \textbf{Drebin}~\cite{arp2014drebin}: This approach first extracts features like permissions, intents, code strings, and API calls from APKs using static analysis, and then trains an SVM classifier to detect malware.
 \item \textbf{RAMDA}~\cite{li2021robust}: This method is a state-of-the-art approach aimed at improving detectors' robustness against adversarial attacks. Specifically, it introduces a Variational Autoencoder (VAE) to learn a compact representation from its input features (\ie intents, permissions, and API calls) and then trains a binary classifier for malware detection.
\end{itemize}

\subsection{Robustness Enhancement (RQ1)}
\label{sec:robust}

\bulletpoint{Settings}
In this RQ, we investigate whether \codename can effectively enhance its robustness against adversarial attacks compared to state-of-the-art detectors.
Similar to previous work~\cite{li2023black,liu2024unraveling}, we measure the resilience of these detectors against adversarial attacks using two metrics: (a) Attack Success Rate (ASR), which indicates the percentage of adversarial examples that successfully evade detection, and (b) Average Perturbation Ratio (APR), which quantifies the average percentage of perturbed edges in the graph representation.
Note that we only report APR on \codename and its graph-based competitors, as it is not comparable with approaches that use different feature representations.
For a fair comparison, we conduct this experiment on the whole dataset from 2016 to 2020 and set the maximum number of iterations to 100 for all models.

To comprehensively explore the robustness of \codename, we implement a representative attack algorithm, Integrated Gradient Guided JSMA (IG-JSMA) attack~\cite{wu2019adversarial}, under two distinct scenarios: white-box and black-box. JSMA has been widely employed to craft adversarial examples (AEs) for evading Android malware detectors \cite{wu2019malscan, chen2019android, liu2024unraveling}. Specifically, it perturbs the most influential features based on indicative forward derivatives to create AEs. 
Additionally, we adopt previous strategies~\cite{chen2019android,li2023black} to ensure the generated adversarial examples can be repacked into APKs.
For feature-vector-based approaches like Drebin, we follow \cite{chen2019android} by constraining modifications to focus on vector bits that are 0, changing 0s to 1s. This ensures that all required permissions or functions remain unchanged, preserving functionality.
For graph-based approaches like \codename, we introduce edges that do not affect APK functionality, in line with the technique described in~\cite{li2023black}.



\begin{figure}[t]
 \centering
 \includegraphics[width=0.95\linewidth]{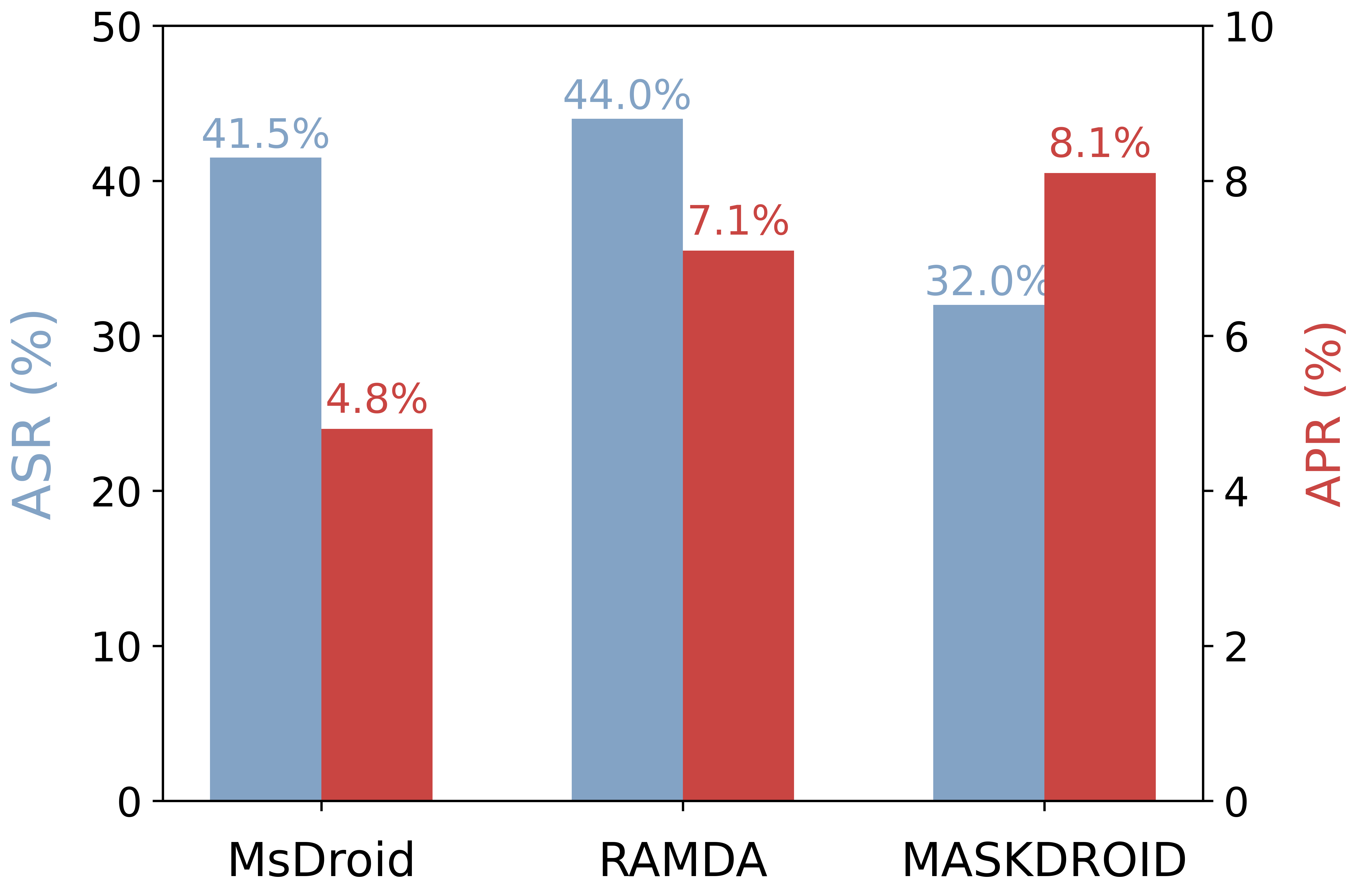}
  \vspace{-10pt}
 \caption{Robustness Evaluation against White-Box Adversarial Attacks (mask rate $\gamma$= 0.8).}
 \vspace{-5pt}
 \label{fig:asr_apr}
\end{figure}

\subsubsection{White-Box Attack Defense}
\label{sec:whitebox}
White-box adversarial attacks occur when attackers possess full knowledge of the victim model, including training data, model structure, parameters, gradient information, and prediction results~\cite{goodfellow2014}.
In this scenario, we exclude Mamadroid~\cite{mamadroid}, Malscan~\cite{wu2019malscan}, and Drebin~\cite{arp2014drebin}, as they utilize traditional machine learning methods rather than deep neural networks, preventing us from calculating their gradient information.


Figure~\ref{fig:asr_apr} illustrates the results of MsDroid~\cite{he2022msdroid}, RAMDA~\cite{li2021robust}, and \codename against white-box adversarial attacks.
From this figure, we observe that \codename consistently outperforms its competitors in terms of both ASR (32.0\%) and APR (8.1\%).
Specifically, the lower ASR indicates that \codename is more resilient against adversarial attacks than MsDroid and RAMDA.
Additionally, the higher APR suggests that \codename requires perturbing more edges to evade detection, meaning adversaries need to exert more effort to craft adversarial examples.
This can be attributed to the stable representations learned by \codename by reconstructing the whole graph from a small portion of nodes, which enhances its robustness against adversarial attacks.

   


\begin{table}[t]
    \centering
    \caption{Robustness Evaluation against Black-Box Adversarial Attacks (mask rate $\gamma$= 0.8).}
     \vspace{-10pt}
    \begin{adjustbox}{width=0.82\linewidth, center}
        \begin{tabular}{@{}c|c|c|c@{}}
            \toprule
            \multicolumn{1}{l|}{\textbf{Detectors}} & \textbf{Malscan} & \textbf{MamaDroid}& \textbf{Drebin}    \\ \midrule
            \textbf{ASR}       & 98.5\%           & 69.0\%            & 100\%              \\ \midrule
            \textbf{APR}       & -                & -                 & -                  \\ \midrule
            \multicolumn{1}{l|}{\textbf{Detectors}} & \textbf{MsDroid} & \textbf{RAMDA} & \textbf{\codename} \\ \midrule
            \textbf{ASR}       & 13.2\%           & 19.2\%            & 19.1\%             \\ \midrule
            \textbf{APR}       & 0.5\%            & 1.5\%             & 10.1\%             \\ \bottomrule
            \end{tabular}
    \end{adjustbox}
     \vspace{-0.3cm}
    \label{tab:robustness}
\end{table}

\subsubsection{Black-Box Attack Defense}
\label{sec:blackbox}
In real-world scenarios, attackers do not always have access to the detailed structures and parameters of the malware detection systems and are often limited to certain knowledge, such as the training dataset and classification results~\cite{hu2022generating, li2019adversarial, li2023black}.
In this context, the typical attack strategy is to distill a substitution model that mimics the behavior of the original black-box model, and then calculate gradients based on the substitution model to simulate a white-box attack.
We follow the same strategy to conduct a black-box attack on the substitution model.
Specifically, we use an MLP to mimic the learning models of MamaDroid~\cite{mamadroid}, MalScan~\cite{wu2019malscan}, Drebin~\cite{arp2014drebin} and RAMDA~\cite{li2021robust}.
For the model of MsDroid~\cite{he2022msdroid}, we use a two-layer GNN encoder followed by an MLP classifier to mimic its behavior.
To train the substitution models, we use the labels predicted by the target black-box models.
For example, we first collect the predictions of MamaDroid on the training set, then train an MLP to mimic MamaDroid's behavior using the collected predictions as labels.

Table~\ref{tab:robustness} presents the results of \codename and its baselines against black-box adversarial attacks.
Note that we exclude the APR for MamaDroid, MalScan, and Drebin, as they do not use graph representations.
From the table, we find that Drebin~\cite{arp2014drebin} is the most vulnerable to black-box adversarial attacks, with an ASR of 100\%.
This is because Drebin is based on binary features, which are easy to manipulate.
We know that Malscan~\cite{wu2019malscan} and MamaDroid~\cite{mamadroid} extract features from the FCGs, while MamaDroid is more robust than Malscan, with an ASR of 69.0\%.
This increased robustness is because MamaDroid abstracts the FCGs according to the package or family level of function names, making it more difficult to perturb, aligning with previous findings~\cite{gao2024comprehensive}.
MsDroid~\cite{he2022msdroid} and RAMDA~\cite{li2021robust} achieve ASRs of 13.2\% and 19.2\%, respectively, which are lower or comparable to \codename.
While \codename achieves the highest APR (10.1\%), indicating it is very difficult to attack in real-world scenarios.
Specifically, a higher APR indicates that more iterations are needed and more edges must be modified to craft adversarial examples (AEs).
Given the large app graph size and the constraint that AEs can be repacked into APKs, finding a usable edge is not trivial, making the attack ineffective. 




\begin{center}
 \begin{conclusionbox}
 \textit{Result 1:}
 Compared with state-of-the-art Android malware detection approaches, \textit{\codename} enhances robustness against adversarial attacks in both white-box and black-box scenarios. Notably, in the white-box attack, \textit{\codename} reduces the attack success rate by 9.5\% against the second-best baseline.
 \end{conclusionbox}
\end{center}

\subsection{Effectiveness Comparison (RQ2)}
\label{sec:effect}
\bulletpoint{Settings}
Having verified \codename's robustness against adversarial attacks, we now investigate whether this robustness comes at the expense of detection effectiveness.
To measure the detection effectiveness of \codename and its baselines, we evaluate their performance on the testing set using standard metrics, including precision, recall, F1-score, and accuracy, following the standard practice in Android malware detection~\cite{wu2019malscan,he2022msdroid}.
Specifically, precision and recall measures correctly detect malware against all detected malware and all actual malware, respectively.
The F1-score, calculated as $2 \times \frac{precision \times recall}{precision + recall}$, represents the balance between precision and recall.
Accuracy is the ratio of correctly classified apps to the total number of analyzed apps.
This experiment is conducted on the dataset from 2016 to 2020.
Additionally, temporal bias is widely recognized as a key factor influencing the effectiveness of malware detectors~\cite{liu2024unraveling}.
We also explore its impact on \codename.
Specifically, we train the model using data collected from 2016 to 2019 and evaluate it on data from 2020.

 





\begin{table}[t]
    \centering
    \caption{The detection effectiveness of \codename and its baselines on the dataset from 2016 to 2020.}
     \vspace{-10pt}
    \begin{adjustbox}{width=0.85\linewidth, center}
   \begin{tabular}{@{}l|c|c|c|c@{}}
  \toprule
  \textbf{Detectors} & \textbf{Precision} & \textbf{Recall} & \textbf{F1} & \textbf{Accuracy} \\ \midrule
  \textbf{Malscan}   & 0.811    & 0.805 & 0.808  & 0.962   \\ \midrule
  \textbf{MamaDroid} & 0.922    & 0.768 & 0.838  & 0.970   \\ \midrule
  \textbf{Drebin}    & 0.763    & 0.712 & 0.736  & 0.949   \\ \midrule
  \textbf{MsDroid}   & 0.582    & 0.615 & 0.598  & 0.917   \\ \midrule
  \textbf{RAMDA} & 0.821    & 0.800 & 0.811  & 0.962   \\ \midrule
  \textbf{\codename} & 0.709    & 0.876 & 0.783  & 0.951   \\ \bottomrule
  \end{tabular}
    \end{adjustbox}
     \vspace{-10pt}
    \label{tab:detection}
\end{table}


Table~\ref{tab:detection} demonstrates the detection effectiveness of \codename and its baselines.
From the results, we observe that \codename achieves a precision of 0.709, recall of 0.876, F1-Score of 0.783, and accuracy of 0.951.
While \codename's F1-Score is lower than some baselines, such as MalScan~\cite{wu2019malscan} and MamaDroid~\cite{mamadroid}, it is important to note that \codename's F1-Score is still competitive.
We can see that \codename achieves the highest recall among all baselines, indicating its superior ability to detect Android malware.
This phenomenon is in line with the design of \codename, which aims to enhance the model's capability to counteract adversarial attacks.
By training the model to learn the underlying semantics and identify more malicious signals in the graph, \codename is more likely to detect a greater number of malware samples and adversarial examples.
However, this also leads to the misclassification of some benign apps as malware, resulting in lower precision compared to several baselines.
Given the trade-off between detection robustness and effectiveness, we believe that \codename's performance is satisfactory and competitive with existing Android malware detectors.

Table~\ref{tab:concept_drift} presents the detection effectiveness of \codename and its baseline models in the presence of temporal bias.
We observe that \codename achieves superior results in terms of F1-score, demonstrating its ability to detect malware even when faced with temporal bias.
This can be attributed to \codename’s focus on learning high-quality representations that encode malicious patterns, which remain stable and robust over time.


\begin{center}
 \begin{conclusionbox}
 \textit{Result 2:} 
\codename achieves detection effectiveness comparable to existing Android malware detectors in both same-time distribution scenarios and situations involving temporal bias.
 \end{conclusionbox}
\end{center}

\begin{table}[t]
    \centering
    \caption{The detection effectiveness of \codename and its baselines on the temporal biased dataset.}
     \vspace{-10pt}
    \begin{adjustbox}{width=0.85\linewidth, center}
   \begin{tabular}{@{}l|c|c|c|c@{}}
  \toprule
  \textbf{Detectors} & \textbf{Precision} & \textbf{Recall} & \textbf{F1} & \textbf{Accuracy} \\ \midrule
  \textbf{Malscan}   & 0.650 & 0.372 & 0.473 & 0.916  \\ \midrule
  \textbf{MamaDroid} & 0.934 & 0.272 & 0.421 & 0.924    \\ \midrule
  \textbf{Drebin}    & 0.743 & 0.466 & 0.573 & 0.929    \\ \midrule
  \textbf{MsDroid}   & 0.642 & 0.211 & 0.317 & 0.908  \\ \midrule
  \textbf{RAMDA}     & 0.531 & 0.562 & 0.546 & 0.905   \\ \midrule
  \textbf{\codename} & 0.541 & 0.630 & 0.582 & 0.908   \\ \bottomrule
  \end{tabular}
    \end{adjustbox}
     \vspace{-10pt}
    \label{tab:concept_drift}
\end{table}

\subsection{Ablation Study on \codename (RQ3)}
\label{sec:ablation}
In this section, we investigate the impact of different design choices on the performance of \codename.
Specifically, we conduct ablation experiments on various components (\ie reconstruction module, contrastive module, and the masking mechanism) to explore how they contribute to \codename's robustness and effectiveness in Android malware detection.
 
\subsubsection{Effect of Reconstruction/Contrastive Modules}
\label{sec:two_module}
\begin{figure}[t]
 \centering
 \includegraphics[width=0.86\linewidth]{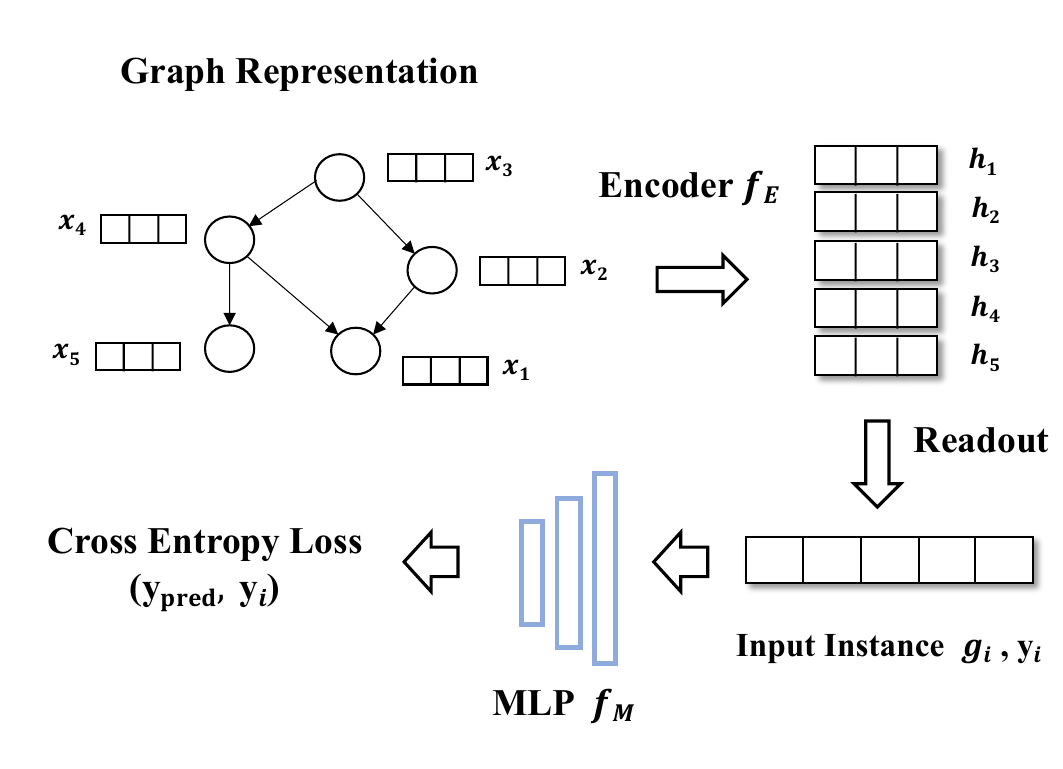}
  \vspace{-0.2cm}
 \caption{Framework of the model \codename-cr that disables both the contrastive module and the reconstruction module from\codename. The input graph goes through a two-layer GNN encoder, proceeds with a readout layer, and is then fed into an MLP classifier.}
  \vspace{-0.3cm}
 \label{fig:baseline}
 
\end{figure}

To clarify our description, we first introduce the terminology used in this section.
\textit{\codename-cr} refers to a version of \codename where both the contrastive and reconstruction modules have been removed, as illustrated in Figure~\ref{fig:baseline}.
\textit{\codename-c} means replacing the contrastive module in Figure~\ref{fig:framework} with an MLP as the binary classifier.
\textit{\codename-r} denotes disabling the reconstruction mechanism in Figure~\ref{fig:framework}, and feeding the readout from encoder $f_E$ directly into the contrastive module.

Table~\ref{tab:two_module_perf} presents the detection effectiveness of \codename and its ablated versions.
As shown, \codename achieves the highest performance with an F1-Score of 82.4\%.
When the contrastive module (\codename-c) or the reconstruction module (\codename-r) is removed, the F1-Score drops to 77.4\% and 79.9\%, respectively.
This indicates that both the reconstruction and contrastive modules are essential for \codename to achieve optimal performance in malware detection.
Interestingly, the model without either module (\codename-cr) still achieves a relatively high F1-Score, surpassing \codename-c and \codename-r, which suggests that the two modules complement each other to enhance the model's performance. 
We can also observe a clear trend in precision and recall, that is, \codename has lower precision and higher recall compared to when two modules are removed. This is because \codename uses reconstruction and contrastive modules to enhance its ability to capture malicious signals, significantly improving recall (the probability of detecting more malware). Although this introduces several false positives, the overall F1-score continues to improve.

\begin{figure}[t]
 \centering
 \includegraphics[width=0.97\linewidth]{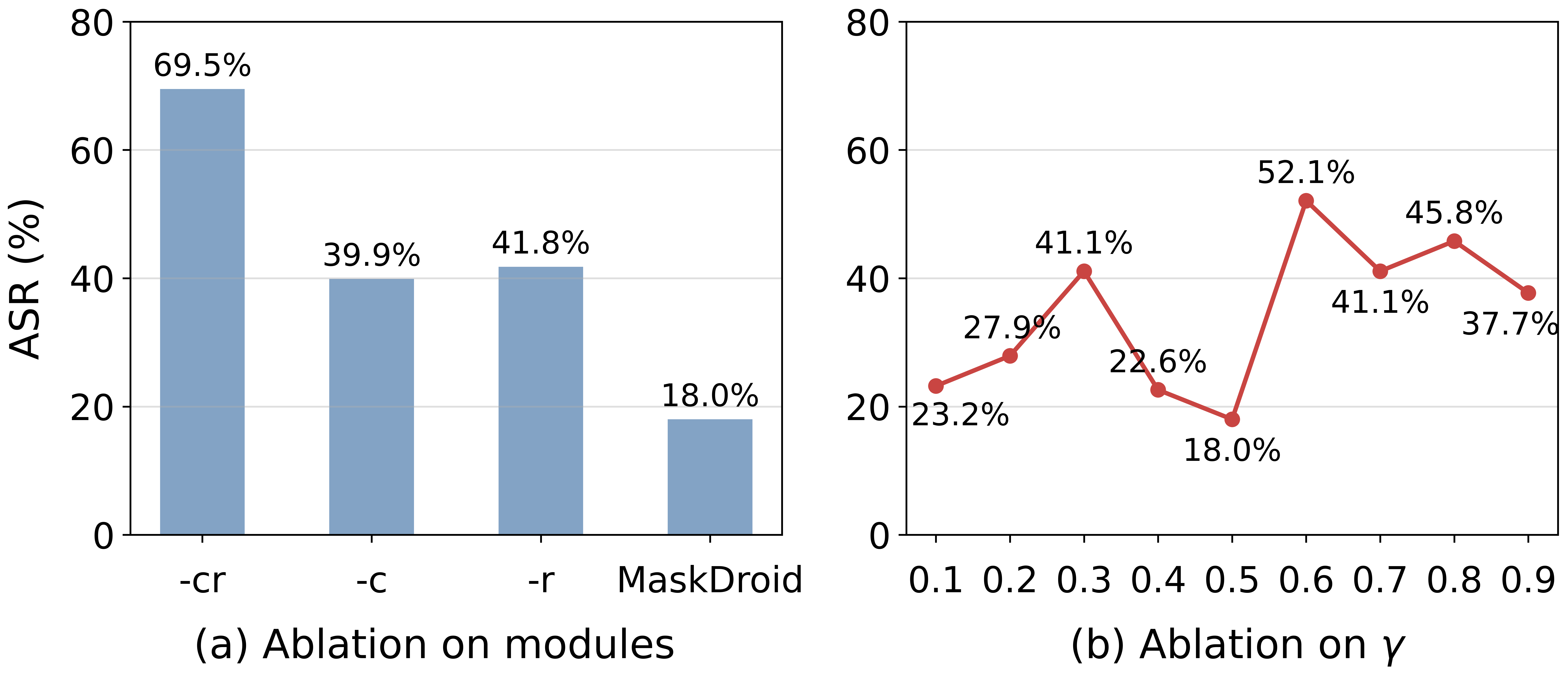}
  \vspace{-0.2cm}
 \caption{Ablation study on reconstruction/contrastive modules and mask rate $\gamma$ for \codename's robustness against white-box adversarial attacks. The left subfigure presents the results of the reconstruction/contrastive modules, while the right subfigure illustrates the impact of the mask rate $\gamma$.}
  \vspace{-0.3cm}
 \label{fig:ablation}
\end{figure}

We further evaluate the robustness of \codename and its ablated versions against adversarial attacks.
Given that white-box adversarial attacks are more effective than black-box attacks, our focus in this ablation study is on the former.
The left part in Figure~\ref{fig:ablation} demonstrates the impact of the reconstruction and contrastive modules on the model's robustness against the white-box adversarial attack.
The results show that \codename significantly lowers the Attack Success Rate (ASR) compared to \codename-c, \codename-r, and \codename-cr, indicating its superior robustness against adversarial attacks.
Comparing \codename-c and \codename-r with \codename-cr, we note that both \codename-c and \codename-r exhibit a lower ASR, suggesting that the contrastive and reconstruction modules are crucial for enhancing the model's robustness.


\begin{table}[t]
 \centering
 \caption{Ablation study on reconstruction and contrastive modules for Android malware detection performance.}
  \vspace{-10pt}
 \begin{adjustbox}{width=0.84\linewidth, center}
\begin{tabular}{@{}l|c|c|c|c@{}}
\toprule
\textbf{Models} & {\textbf{Precision}} & {\textbf{Recall}} & \textbf{F1} & \textbf{Accuracy} \\ \midrule
\textbf{\codename-cr} & {0.918}& {0.730} & 0.813 & 0.965 \\ \midrule
\textbf{\codename-c}& {0.886}& {0.688} & 0.774 & 0.958 \\ \midrule
\textbf{\codename-r}& {0.896}& {0.720} & 0.799 & 0.962 \\ \midrule
\textbf{\codename}& {0.772}& {0.883} & 0.824 & 0.961 \\ \bottomrule
\end{tabular}
 \end{adjustbox}
  \vspace{-0.3cm}
 \label{tab:two_module_perf}
\end{table}

\subsubsection{Effect of mask rate $\gamma$}
\label{sec:maskrate}
The masking mechanism is one of our key designs to encourage \codename to learn a more holistic representation of the input graph.
Selecting an appropriate mask rate $\gamma$ is crucial for our model's performance.
In this section, we vary the mask rate (\ie 0.2, 0.5, 0.8, 0.9) to examine its influence on the detection effectiveness and robustness of \codename.

Table~\ref{tab:maskrate} presents the changes in detection performance as the mask rate $\gamma$ varies.
We observe that \codename's detection effectiveness is not overly sensitive to the mask rate.
Specifically, when the mask rate increases from 0.2 to 0.9, the F1-Score falls slightly from 0.789 to 0.824.
One potential explanation is that \codename's contrastive module can effectively learn the differences between the benign and malicious samples, even when the input graph is partially masked.
Delving deeper into the robustness of \codename against adversarial attacks, we find that the model's performance is more sensitive to the mask rate.
As shown in the right part of Figure~\ref{fig:ablation}, the ASR is lowest (0.180) when the mask rate is 0.5, and increases to 0.458 when the mask rate is 0.8.
This may be because a higher mask rate introduces more noise into the input graph, making it more challenging for the model to learn the underlying patterns.
Therefore, selecting an appropriate mask rate is crucial for \codename to achieve optimal performance in both malware detection and defense against adversarial attacks.
We leave how to automatically determine the mask rate as future work.

\begin{table}[t]
\renewcommand{\arraystretch}{0.8}
\centering
\caption{Ablation study on mask rate $\gamma$ for Android malware detection performance.}
 \vspace{-0.1cm}
\begin{adjustbox}{width=0.8\linewidth, center}
 \begin{tabular}{@{}c|c|c|c|c@{}}
 \toprule
 \textbf{Mask Rate} & {\textbf{Precision}} & {\textbf{Recall}} & \textbf{F1} & \textbf{Accuracy} \\ \midrule
 \textbf{$\gamma = 0.1$} & {0.685}  & {0.888}   & 0.773  & 0.947 \\ \midrule
 \textbf{$\gamma = 0.2$} & {0.720}  & {0.874}   & 0.789  & 0.952 \\ \midrule
 \textbf{$\gamma = 0.3$} & {0.744}  & {0.851}   & 0.794  & 0.955 \\ \midrule
 \textbf{$\gamma = 0.4$} & {0.751}  & {0.884}   & 0.812  & 0.958 \\ \midrule
 \textbf{$\gamma = 0.5$} & {0.772}  & {0.884}   & 0.824  & 0.961 \\ \midrule
 \textbf{$\gamma = 0.6$} & {0.756}  & {0.865}   & 0.807  & 0.958 \\ \midrule
 \textbf{$\gamma = 0.7$} & {0.757}  & {0.884}   & 0.816  & 0.959 \\ \midrule
 \textbf{$\gamma = 0.8$} & {0.728}  & {0.897}   & 0.804  & 0.955 \\ \midrule
 \textbf{$\gamma = 0.9$} & {0.724}  & {0.869}   & 0.790  & 0.952 \\ \bottomrule
 \end{tabular}
\end{adjustbox}
 \vspace{-0.3cm}
\label{tab:maskrate}
\end{table}

\subsubsection{Visualization of Representations}
The main contribution of \codename is to learn high-quality representations that grasp the holistic understanding of app behaviors and derive a clear decision boundary to achieve impressive discriminative power for both malware and adversarial examples.
To further understand the internal representations learned by \codename, we visualize the latent representations obtained by \codename and its variant \codename-cr, which disables the reconstruction and contrastive modules.
We adopt the t-SNE technique to project the graph representation for each input sample onto a 2D space~\cite{van2008visualizing}.

\begin{figure}[t]
	\centering
	\subcaptionbox{\codename \label{fig:rep_maskdroid}}{
	  \vspace{-6pt}
		\includegraphics[width=0.43\linewidth]{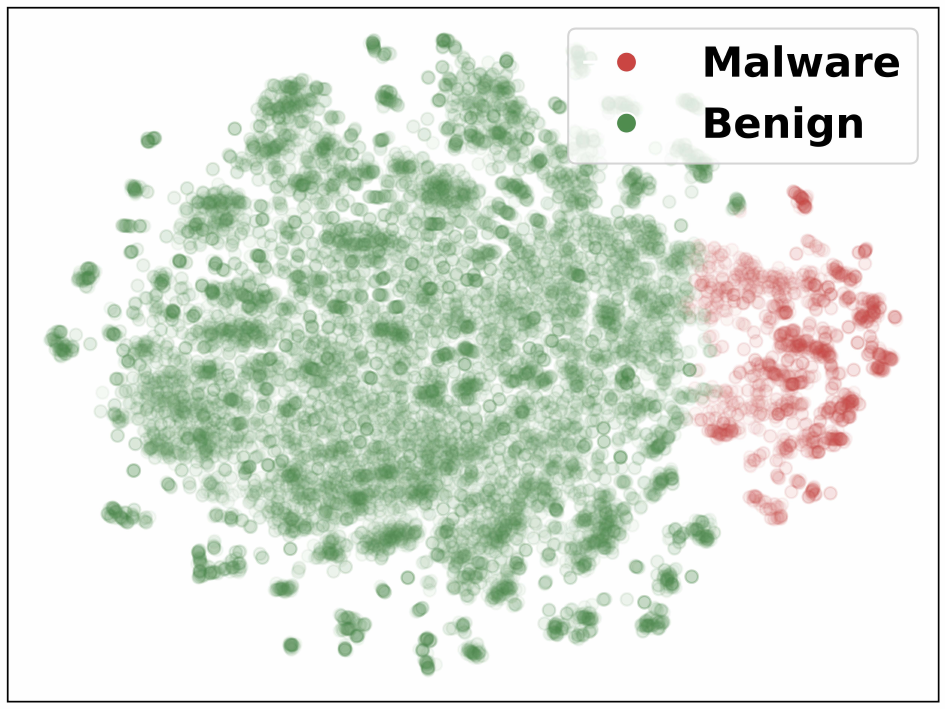}}
	\subcaptionbox{\codename-cr\label{fig:rep_baseline}}{
	  \vspace{-6pt}
		\includegraphics[width=0.43\linewidth]{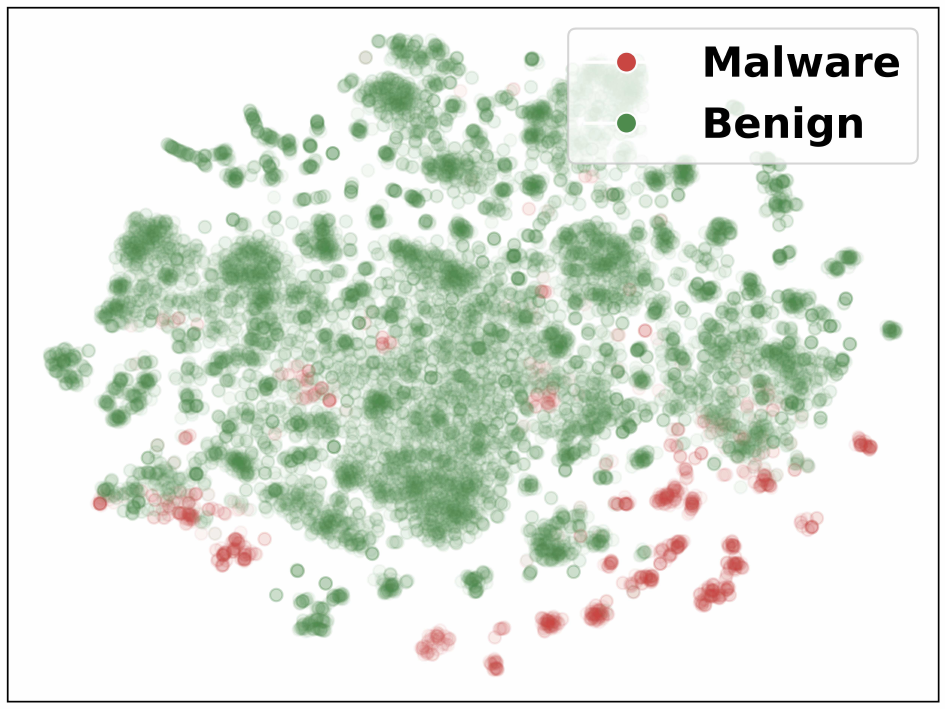}}
	 \vspace{-8pt}	
	\caption{
 A more compressed representation learned by \codename compared to \codename-cr, which disables the reconstruction module and the contrastive module.}
	\label{fig:vis}
	\vspace{-0.3cm}
\end{figure}

Figure~\ref{fig:vis} illustrates the representations learned by \codename and \codename-cr.
In the figure, the green points represent the representations of benign samples, while the red points represent those of malicious samples.
We can see that the latent representations learned by \codename are more compact than those learned by \codename-cr.
Also note that \codename achieves a much clearer decision boundary, where the benign and malicious samples are more distinctly separated.
The compact representations and clear boundary not only enable the model to excel in classification tasks but also significantly increase the difficulty of adversarial attacks, thus resulting in its enhanced robustness~\cite{fan2021does}.
This can be attributed to the ability of the reconstruction and contrastive modules to learn the underlying patterns in the input graph and differentiate different types of samples.




\begin{center}
 \begin{conclusionbox}
 \textit{Result 3:} 
 All of our choices in designing \codename, including the reconstruction and contrastive modules, as well as the mask rate $\gamma$, are crucial for enhancing the model's effectiveness and robustness in malware detection.
 \end{conclusionbox}
\end{center}

\subsection{Efficiency Evaluation (RQ4)}
\label{sec:efficiency}
In addition to the robustness and effectiveness of \codename, efficiency is another critical factor influencing the model's practicality.
In this section, we compare the training costs of \codename with its baseline and variant models to evaluate its efficiency.
For a fair comparison, we exclude MamaDroid~\cite{mamadroid}, Malscan~\cite{wu2019malscan}, and Drebin~\cite{arp2014drebin}, as they utilize traditional machine learning models to extract features and train classifiers, which are not comparable to deep learning models in terms of training cost.
It is important to highlight that when training these detectors, we apply the early stopping strategy to prevent over-fitting.
Additionally, the training process is conducted on the same server with the same configuration as mentioned in Section~\ref{sec:implementation}.

Table~\ref{tab:elapse_time} summarizes the training costs of \codename and its baseline or variant models.
We observe that \codename requires 150 epochs to converge, lasting 2,100 seconds in total.
This is comparable to MsDroid and significantly less than RAMDA's strategy, indicating that \codename does not sacrifice efficiency for robustness.
By further comparing \codename with its variant models (\ie, -cr, -c, -r), we observe that \codename-cr takes the least time to converge, which is expected as it does not involve the reconstruction and contrastive modules.
We also note that \codename requires a comparable amount of time to \codename-r and \codename-c, indicating that the two modules are not simply additive but work in meaningful cooperation to reinforce each other and achieve the goal of robustness.

\begin{center}
 \begin{conclusionbox}
 \textit{Result 4:} 
 \codename achieves a balance between detection robustness and efficiency, demonstrating superior resilience against adversarial attacks while maintaining a moderate training cost compared to its baselines. 
 \end{conclusionbox}
\end{center}

\begin{table}[t]
 \centering
 \caption{Comparison of training costs between \codename and its baseline/variant models (on 2020 data).}
 \vspace{-10pt}
 \begin{adjustbox}{width=0.86\linewidth, center}
 \begin{tabular}{@{}l|cccccc@{}}
\toprule
\multirow{2}{*}{} & \multirow{2}{*}{\textbf{MsDroid}} & \multirow{2}{*}{\textbf{RAMDA}} & \multicolumn{3}{c}{\textbf{\codename}}& \multirow{2}{*}{\textbf{\codename}} \\
&&& \textbf{-cr} & \textbf{-c} & \textbf{-r} &\\ \midrule
\textbf{Epoch(s)} & 186 & 240 & 60 & 128& 115& 150\\
\textbf{Total(s)} & 1,860 & 12,770& 600& 2,560& 8,050& 2,100\\ \bottomrule
\end{tabular}
 \end{adjustbox}
  \vspace{-10pt}
 \label{tab:elapse_time}
\end{table}





\section{Threats to Validity}
\label{sec:threats_validity}
In this section, we discuss the threats to the validity of our study.
First, the effectiveness and robustness of \codename may be influenced by different hyper-parameters used in the neural networks.
To mitigate this threat, we adopt advanced practices from prior studies~\cite{liu2023learning,zengy2022shadewatcher}, employing a grid search to tune and find the optimal hyper-parameters that yield the best performance on validation sets.
We detail all hyper-parameter settings in Section~\ref{sec:setup} for reproducibility and release our evaluation artifacts, including codes and datasets.
Second, when conducting comparison experiments, we utilize open-source implementations of baseline methods.
However, since we have tuned the hyper-parameters of these baselines to suit our datasets and experimental settings, the performance of them may differ from the results originally reported in the literature.
Third, we do not compare \codename's defensive capabilities with other adversarial defense methods, such as adversarial training~\cite{al2018adversarial}.
This is because these methods are orthogonal to our work and can be integrated with \codename to enhance further its robustness against adversarial attacks, which we leave as future work.
At last, following previous studies~\cite{wu2019malscan,chen2019android}, we use the representative attack algorithm JSMA to evaluate \codename’s robustness.
However, \codename’s resilience against new and evolving attack methods remains unexplored, which we leave for future work.
\section{Related Work} 
\label{sec:related_work}

In this section, we begin by reviewing related work on machine learning (ML)-based Android malware detection. Subsequently, we describe the masking mechanism.
Finally, we introduce common adversarial example generation methods and their corresponding defense strategies.

\bulletpoint{ML-based Android Malware Detection}
Machine learning (ML) techniques have been extensively employed to analyze various types of APK features and extract malicious patterns for Android malware detection.
Syntactical features, such as permissions, API calls, code strings, and intents, are commonly used to model app behaviors~\cite{arp2014drebin,wu2021android,li2021robust,aafer2013droidapiminer,peng2012using}.
For example, Xmal~\cite{wu2021android} uses an attention-based MLP to distill malicious signals from the API calls and permissions.
RAMDA~\cite{li2021robust} feeds intents and API calls into an Autoencoder to learn a resilient representation of the APKs for malware detection.
Drebin~\cite{arp2014drebin} first utilizes static analysis to extract features such as permissions, intents, code strings, and API calls and then trains an SVM classifier to identify Android malware.
However, such methods may fail to capture the program semantics of apps, limiting the detection effectiveness.
To mitigate this issue, researchers turn to extracting different app semantics for effective Android malware detection~\cite{xu2018deeprefiner,wu2019malscan,wu2021homdroid,mamadroid,he2022msdroid,karbab2021petadroid,allen2018improving,fan2017dapasa,hou2017hindroid,feng2014apposcopy,mclaughlin2017deep}.
DeepRefiner~\cite{xu2018deeprefiner} and Mclaughlin et al.~\cite{mclaughlin2017deep} represent app bytecodes as texts and images, respectively, applying LSTM and CNN to detect malware.
Malscan~\cite{wu2019malscan} and HomDroid~\cite{wu2021homdroid} treat the function call graphs of apps as social networks and apply the corresponding centrality analysis algorithm to capture malicious patterns for Android malware detection.
Furthermore, MsDroid~\cite{he2022msdroid} describes app-sensitive behaviors with code snippets around sensitive API calls and employs graph neural networks to distill the semantic and structural information to identify malware.

\bulletpoint{Masking Mechanisms} In the field of graph representation learning, masking and reconstructing graph features has proven to be an effective approach for achieving robust learning \cite{hou2022graphmae}. Several works have applied masking mechanisms to improve model robustness in areas including image classification and text classification \cite{PatchGuard, MASKER}. 
For example, PatchGuard \cite{PatchGuard} proposed a masking defense to obscure corrupted features and recover the correct prediction for image classification tasks. MASKER \cite{MASKER} regularizes language models to reconstruct keywords from the remaining words and make low-confidence predictions when there is insufficient context. Inspired by the success of masking mechanisms in improving model robustness, we pioneer and adapt this practice to enhance the robustness of Android malware detection.

\bulletpoint{Adversarial Example Attack}
With ML-based Android malware detection evolving, attackers increasingly seek to evade these detectors by purposefully perturbing the malicious APK samples.
These attacks can be categorized into feature-space attacks~\cite{grosse2017adversarial,shahpasand2019adversarial,hu2022generating,al2018adversarial,li2019adversarial} and problem-space attack~\cite{li2023black,zhao2021structural,pierazzi2020intriguing,chen2019android,li2020adversarial}.
Feature-based attacks target the feature vectors extracted from APKs directly, intending to deceive detectors.
For instance, Hu et al.~\cite{hu2022generating} leverage the capability of generative adversarial networks (GANs)~\cite{goodfellow2020generative} to modify binary feature vectors, aiming to mislead the detectors.
Grosse et al.~\cite{grosse2017adversarial} propose a Jacobian matrix-based method to manipulate the features, allowing malicious samples to escape detection.
In contrast, problem-based attacks endeavor to produce real adversarial malware.
Specifically, HRAT~\cite{zhao2021structural} utilizes reinforcement learning to strategically modify the structure of a malicious app without affecting its original functionality.
Li et al.~\cite{li2023black} combine GAN with a multi-population co-evolution algorithm to add \textit{try-catch} edges for crafting adversarial examples.
Pierazzi et al.~\cite{pierazzi2020intriguing} explore the creation of adversarial malware by utilizing bytecode slices extracted from benign APKs.
Furthermore, Android HIV~\cite{chen2019android} establishes a correspondence between the feature space and problem space, ensuring that any feature perturbations do not compromise the core functionality of the malware.

\bulletpoint{Adversarial Example Defense}
To combat adversarial attacks, one can approach the problem from two distinct angles: by fortifying the data~\cite{pendlebury2019tesseract,goodfellow2014,al2018adversarial,bhagoji2018enhancing,meng2017magnet} or by enhancing the model~\cite{li2021robust,distillation,tramer2017ensemble,dhillon2018stochastic}.
For example, Huang et al.~\cite{al2018adversarial} devise a method to create a diverse set of functionally preserved adversarial examples.
By incorporating these diverse samples into the training process, they aim to bolster the model's resilience against adversarial attacks.
Goodfellow et al.~\cite{goodfellow2014} suggest mitigating adversarial example attacks through model retraining. 
They incorporate the original dataset with newly labeled adversarial malware examples, enhancing the classifier's familiarity with various malware.
Taking a different route, Bhagoji et al.~\cite{bhagoji2018enhancing} employ Principal Component Analysis (PCA) to project all inputs into a lower-dimensional space, thus diminishing the model's susceptibility to adversarial attacks.
Turning to strategies that directly fortify the model itself, RAMDA~\cite{li2021robust} makes use of an autoencoder to derive a compressed representation of APKs, thereby filtering out potential adversarial examples.
Similarly, Papernot et al.~\cite{distillation} adopt defense distillation to mitigate the vulnerability of deep neural networks against minor perturbations.

\section{Conclusion}
In this work, we propose \codename, a novel framework designed to enhance robustness against adversarial attacks while maintaining impressive discriminative power for Android malware detection.
Specifically, we introduce a masking mechanism and force \codename to reconstruct the entire graph using the unmasked part, enabling it to learn stable representations of the input graphs more effectively.
Additionally, \codename incorporates a contrastive module to cluster samples of the same class together while pushing samples of different classes apart, thereby enhancing the model's discriminative power.
Grounded by extensive evaluations, \codename steadily outperforms state-of-the-art (SOTA) baselines on adversarial attack defense tasks and achieves comparable performance on malware detection tasks.
A promising direction for future work would be to extend the exploitation of the mask mechanism to attention masks with semantic information, which could further enhance the model's understanding of malicious behaviors.
Furthermore, since our strategy is applied to graph-based data and is not restricted to specific graph structures, similar methods can be extended to other graph-based datasets, such as control flow graphs, to boost the model's robustness.

\begin{acks}
This research/project is supported by the National Research Foundation, Singapore under its Industry Alignment Fund – Pre-positioning (IAF-PP) Funding Initiative and the NExT project.
Any opinions, findings and conclusions or recommendations expressed in this material are those of the authors and do not reflect the views of National Research Foundation, Singapore.
\end{acks}

\bibliographystyle{ACM-Reference-Format}
\balance
\bibliography{paper}

\end{document}